# Pivot of the Emerging Bipolar Magnetic Region in the Birth of Sigmoidal Solar Active Regions

Ronald L. Moore[1,2], Sanjiv K. Tiwari[3,4], V. Aparna[3,4], Navdeep K. Panesar[3,4], Alphonse C. Sterling[2], and Talwinder Singh[5]

[1]Center for Space Plasma and Aeronomic Research (CSPAR), UAH, Huntsville AL, 35805 USA; ronald.l.moore@nasa.gov
[2]NASA Marshall Space Flight Center, Huntsville, AL 35812, USA
[3]Lockheed Martin Solar Astrophysics Laboratory, 3251 Hanover Street Building 203, Palo Alto, CA 94306, USA; tiwari@lmsal.com
[4]Bay Area Environmental Research Institute, NASA Research Park, Moffett Field, CA 94035, USA
[5]Department of Physics & Astronomy, Georgia State University, Atlanta, GA 30303, USA

## Abstract

We present an augmentation to longstanding evidence from observations and MHD modeling that (1) every solar emerging bipolar magnetic region (BMR) is made by an emerging Ω-loop flux rope, and (2) twist in the flux-rope field makes the emerged field sigmoidal. Using co-temporal full-disk coronal EUV images, magnetograms, and continuum images from Solar Dynamics Observatory (SDO), we found and tracked the emergence of 42 emerging single-BMR sigmoidal active regions (ARs) that have sunspots in both polarity domains. Throughout each AR's emergence, we quantified the emerging BMR's tilt angle to the east-west direction (the x-direction in SDO images) by measuring in the continuum images the tilt angle of the line through the (visually located) two centroids of the BMR's opposite-polarity sunspot clusters. As each AR emerges, it becomes either S-shaped (shows net right-handed magnetic twist) or Z-shaped (shows net left-handed magnetic twist) in the coronal EUV images. Nineteen of the ARs become S-shaped and 23 become Z-shaped. For all 42 ARs, in agreement with published MHD simulations of the emergence of a single-BMR sigmoidal AR from a subsurface twisted flux rope, if the AR becomes S-shaped, the emerging BMR pivots counterclockwise, and if the AR becomes Z-shaped, the emerging BMR pivots clockwise. For our 42 ARs, the pivot amount roughly ranges from 10° to 90° and averages about 35°. Thus, at the onset of the emergence of our average emerging Ω-loop flux rope, the magnetic field's twist pitch angle at the flux rope's top edge is plausibly about 35°.

## 1. Introduction

The entire solar atmosphere above the Sun's photospheric surface – the chromosphere, chromosphere-corona transition region, corona, and solar wind – is permeated with magnetic field. The field evidently primarily comes from magnetic-flux-rope Ω loops that are generated in the convection zone and bubble up through the photosphere (e.g., C. Zwaan 1987; R. Ishikawa, S. Tsuneta, & J. Jurcak 2010; H. C. Spruit 2011; R. F. Stein & A. Nordlund 2012; L. van Driel-Gesztelyi & L. M. Green 2015; P. Charbonneau 2020; R. L. Moore, S. K. Tiwari, N. K. Panesar, & A. C. Sterling 2020; Y. Fan 2021). The field continually evolves via emergence of new Ω loops, movement of the field's feet by convection flows in and below the photosphere, and submergence of field by convection-driven flux cancellation at polarity inversion lines (PILs) in the photospheric magnetic flux (e.g., D. Rabin, R. Moore, & M. J. Hagyard 1984; R. Moore & D. Rabin 1985; C. Zwaan 1987). The evolving magnetic field modulates the Sun's luminosity, results in the mega-Kelvin hot corona and its solar-wind outflow, often explodes to make coronal mass ejections (CMEs), flares, and coronal jets, and continually has myriad smaller explosions all over the Sun (e.g., G. L. Withbroe & R. W. Noyes 1977; G. S. Vaiana & R. Rosner 1978; R. Moore & D. Rabin 1985; R. L. Moore, D. A. Falconer, J. G. Porter, & S. T. Suess 1999; K. Shibata, T. Nakamura, T. Matsumoto, et al. 2007; N.-E. Raouafi, M. K. Georgoulis, D. M. Rust, & P. N. Bernasconi 2010; R. L. Moore, A. C. Sterling, J. W. Cirtain, & D. A. Flaconer 2011, D. E. Innes & L. Teriaca 2013; A. C. Sterling, R. L. Moore, D. A. Falconer, & M. Adams 2015; N. K. Panesar, A. C. Sterling, R. L. Moore, et al. 2019; T. Samanta, H. Tian, V. Yurchyshyn, et al 2019; S. K. Tiwari, N. K. Panesar, R. L. Moore, et al. 2019; N. K. Panesar, S. K. Tiwari, D. Berghmans, et al. 2021). Thus, the emergence of flux-rope Ω loops fuels all solar magnetic activity and consequent space weather.

A newly emerged lone flux-rope Ω loop that has emerged in negligible weak ambient magnetic field is seen in photospheric magnetograms as a bipolar magnetic region (BMR; e.g., Y. M. Wang & N. R. Sheeley 1989; R. L. Moore, S. K. Tiwari, N. K. Panesar, & A. C. Sterling 2020). The BMR is a pair of domains of equal opposite-polarity magnetic flux. They are the two opposite-polarity feet of the Ω loop's two legs. Coronal EUV and coronal X-ray images show the overall 3D form of typical BMRs to be roughly that of the emerged Ω loop's potential-field coronal magnetic arch (e.g., Y. Fan 2021). Viewed vertically from above, the projected shape of the arch in such coronal images is roughly that of an ellipse that has its major axis through the centroids of the BMR's two flux domains (e.g., R. L. Moore, S. K. Tiwari, N. K. Panesar, & A. C. Sterling 2020). A measure of the size of a BMR is the distance D between the two centroids. BMRs range in size from as small as a granule ($D \sim 10^3$ km; R. Ishikawa, S. Tsuneta, & J. Jurcak 2010) to as large as the largest convection cells (giant cells) in MHD simulations of the convection zone ($D \sim 2 \times 10^5$ km; Y. M. Wang & N. R. Sheeley 1989; M. S. Miesch, A. S. Brun, M. L. DeRosa, J. Toomre 2008; R. L. Moore, S. K. Tiwari, N. K. Panesar, & A. C. Sterling 2020).

If the magnetic flux $\Phi$ of the Ω loop is less than $\sim 10^{20}$ Mx, the BMR has no sunspots or sunspot pores (L. van Driel-Gesztelyi & L. M. Green 2015). For $10^{20} < \Phi < 5 \times 10^{21}$ Mx, the BMR has sunspot pores but no full-fledged sunspots with penumbra. For $\Phi > 5 \times 10^{21}$ Mx, the BMR has full-fledged sunspots. Each emerged or emerging BMR that has pores or sunspots is a solar active region (AR) and is given a chronological AR number by NOAA (National Oceanic and Atmospheric Administration).

The great majority of BMRs that are large enough to have a NOAA AR number are directed roughly east-west: the BMR's flux of one polarity leads the opposite-polarity flux in the direction of the Sun's rotation, and the tilt of the BMR's major axis – clockwise or counterclockwise – from heliographic east-west is less than 45° (R. F. Howard 1991). Throughout an eleven-year sunspot cycle, the great majority of NOAA-numbered BMRs in the northern hemisphere have the same leading polarity, the great majority of NOAA-numbered BMRs in the southern hemisphere have the opposite leading polarity, and these polarities are the opposite throughout the next cycle. This hemispheric rule for active-region BMRs is named the Hale-Nicholson law (H. Zirin 1988). The Hale-Nicholson law suggests that the Ω loops for active-region BMRs are from flux-rope Ω-loop stitches that bubble up from east-west bands of magnetic field that are generated at the bottom of the convection zone by the Sun's global dynamo process, have opposite direction in opposite hemispheres during each eleven year sunspot cycle, and reverse direction from cycle to cycle (e.g., H. W. Babcock, 1961; R. L. Moore, J. W. Certain, & A. C. Sterling 2016; R. L. Moore, S. K. Tiwari, N. K. Panesar, & A. C. Sterling 2020; P. Charbonneau 2020; Y. Fan 2021).

In coronal EUV and coronal X-ray images, instead of having roughly the overall shape of the zero-twist potential field, for a minority of single-BMR active regions, the shape is greatly deformed from the potential-field shape. For such active regions, the observed sigmoid shape of the coronal field shows that the field has definite overall right-handed or left-handed twist. For right-handed twist (positive helicity), the sigmoid has the shape of an S; for left-handed twist (negative helicity), the sigmoid has the shape of a Z (or of a backward S) (e.g., D. M. Rust & A. Kumar 1996; L. M. Green, B. Kleim, T. Torok, L. van Driel-Gesztelyi, & G. D. R. Attrill 2007). Markedly sigmoidal active regions are much more prone to explode to make flares and CMEs than are non-sigmoidal active regions (R. C. Canfield, H. S. Hudson, & D. E, McKenzie 1999). The present paper concerns the birth of single-BMR sigmoidal active regions.

It is commonly supposed that the magnetic field in the Ω-loop flux rope that emerges to make an obviously sigmoidal single-BMR active region is more twisted about the flux rope's center line than in a flux rope that emerges to make a single-BMR active region for which the coronal field has a more nearly potential-field shape. In MHD simulations of the Ω-loop emergence of a markedly twisted flux rope from below the photosphere into the corona, the emerged coronal magnetic field becomes markedly sigmoidal (e.g., S. E. Gibson, Y. Fan, C. Mandrini, G. Fisher, & P. Demoulin 2004; W. Manchester, T. Gombosi, D. De Zeeuw, & Y. Fan 2004; Fan, Y. 2009; Z. Liu, C. Jiang, X. Feng, P. Zuo, & Y. Wang 2023). In these simulations, the horizontal line through the two centroids of the two opposite-polarity flux domains of the emerging BMR at first is skewed from the horizontal direction of the initial subsurface straight horizontal twisted flux rope, and then gradually pivots to become aligned with the initial flux rope's direction. If the initial flux rope has right-handed twist, the pivot is counterclockwise and the emerged coronal field becomes S-shaped. If the initial flux rope has left-handed twist, the pivot is clockwise and the emerged coronal field becomes Z-shaped.

This paper reports certain observed aspects of the birth of each of forty-two randomly selected single-BMR sigmoidal active regions. Each active region begins emerging well on the disk and becomes sigmoidal as it emerges. The coronal sigmoid is S-shaped for nineteen of the active regions (thirteen of which are in the southern hemisphere) and Z-shaped for twenty-three (nineteen of which are in the northern hemisphere). Our main finding is the following. As each S-shaped sigmoidal active region emerges, the emerging BMR pivots counterclockwise, and as each Z-shaped sigmoidal active region emerges, the emerging BMR pivots clockwise. That this finding is expected from MHD simulations of the Ω-loop emergence of a twisted flux rope into the solar atmosphere has two interrelated implications. First, it adds credence to the underlying premise of MHD simulations of the emergence of single-BMR sigmoidal active regions: the magnetic field that emerges to become a sigmoidal active region is initially a twisted horizontal flux rope below the photosphere. Second, it also adds credence to the tenet that every emerging BMR is the signature of an emerging flux-rope Ω loop.

## 2. Data and Methods

The data we used are primarily from Solar Dynamics Observatory (SDO; W. D. Pesnell, B. J. Thompson, & P. C. Chamberlin 2012). We used full-disk coronal EUV images from SDO's Atmospheric Imaging Assembly (AIA; J. R. Lemen, A. M. Title, D. J. Akin, et al 2012) and full-disk magnetograms from SDO's Helioseismic and Magnetic Imager (HMI; P. H. Scherrer, J. Schou, R. I. Bush, et al 2012). AIA coronal EUV images have 0.6" pixels and 12 s cadence. HMI magnetograms have 0.5" pixels and 45 s cadence. From Solar Monitor (https://www.solarmonitor.org), we obtained the solar latitude and NOAA AR number for each selected AR.

Using Helioviewer (https://helioviewer.org), we searched SDO full-disk images and magnetograms from 2011 September through 2024 March for on-disk single-BMR sigmoidal ARs that can each be followed back to the start of its emergence in HMI magnetograms. We selected only ARs that become noticeably overall sigmoidal (S-shaped or Z-shaped) in images from the AIA 335 Å channel as they emerge. The AIA 335 Å channel predominantly images Fe XVI emission from plasma at temperatures near $10^{6.4}$ K. Of the images from the six AIA EUV channels that image emission of solar plasma at temperatures near or hotter than $10^{5.8}$ K, we think the 335 Å images best show whether an AR's overall coronal magnetic-field shape is noticeably sigmoidal. Our search yielded forty-two single-BMR AR's that in AIA 335 Å images each becomes noticeably sigmoidal as it emerges. Those ARs are listed in Table 1.

For each AR, we measure the BMR's tilt at many times (usually at least every six hours) through nearly all of the BMR's emergence. (The BMR's tilt is the BMR's angle to solar east-west in the plane of the sky, i.e., the BMR's angle to the direction orthogonal to solar north-south through disk center, the direction of the x axis in SDO full-disk images and magnetograms.) The first measurement is within an hour of when the BMR first shows in HMI continuum images at least one sunspot or sunspot pore in each of its two opposite-polarity domains. At each measurement time, for each of the BMR's two opposite-polarity clusters of sunspots and pores, we get from Helioviewer the (x, y) coordinates of the point (i.e., pixel) that we visually judge to be the centroid of the cluster in the HMI continuum image. From the (x, y) coordinates of the centroids of the two clusters, we calculate the tilt angle of the straight line through the two centroid points. We repeat that procedure three times at each measurement time. We take the mean of the obtained three tilt-angle values to be the BMR's measured tilt angle, and take the uncertainty in the measured tilt angle to be the mean's standard uncertainty calculated from the three measured values. The time profile of the BMR's tilt through the BMR's emergence displays the BMR's pivot direction and pivot amount. With increasing time, the profile falls for clockwise pivot and rises for counterclockwise pivot. The pivot amount is the angular difference between the BMR's first and last measured tilt.

From HMI magnetograms of each AR, we obtain the time profile of the emerging AR's magnetic flux by measuring the AR's unsigned magnetic flux at 15-minute time steps through all or nearly all of the AR's emergence. That profile shows when the AR's emergence is finished or nearly finished. The measured flux for these time profiles is the total unsigned flux above 40 G in a field of view that covers the AR and some surrounding area during the AR's emergence, e.g., the field of view in Figures 1 – 4.

## 3. Results

### *3.1. Four Example Emerging Single-BMR Sigmoidal Active Regions*

It is established that most active regions in the Sun's northern hemisphere have left-handed net magnetic twist, and most active regions in the southern hemisphere have right-handed magnetic twist (e.g., Pevtsov et al 1995). Hence, it is no surprise that of our randomly selected 42 emerging sigmoidal active regions, most of the Z-shaped ones are in the northern hemisphere, and most of the S-shaped ones are in the southern hemisphere. Of the 42, 23 are Z-shaped, and 19 are S-shaped. Nineteen of the 23 Z-shaped ARs are in the northern hemisphere, and 13 of the 19 S-shaped ARs are in the southern hemisphere.

In the following four subsections, we present four example emerging single-BMR sigmoidal ARs: one that emerges in the southern hemisphere and becomes S-shaped, one that emerges in the northern hemisphere and becomes S-shaped, one that emerges in the northern hemisphere and becomes Z-shaped, and one that emerges in the southern hemisphere and becomes Z-shaped. For both S-shaped ARs, the emerging BMR pivots counterclockwise, and for both Z-shaped ARs, the emerging BMR pivots clockwise.

#### *3.1.1. Example Emerging S-shaped AR in the Southern Hemisphere*

Our example emerging S-shaped AR in the southern hemisphere is shown in Figure 1 and its animation. The AR is number 13 (AR 11680) in Table 1. It emerges around latitude S28° in 2013 February.

Figurer 1 and its animation show that the BMR pivots counterclockwise as it emerges and the AR becomes S-shaped in AIA 335 Å images, and that the BMR's two opposite-polarity sunspot clusters spread apart from each other as the BMR emerges. In the bottom panel of Figure 1, the time plots of the BMR's tilt and magnetic flux together show that the BMR completes its pivot after it completes nearly all of its emergence. The BMR's first measured tilt is at 17:00 UT on February 24 and its last measured tilt is at 00:00 UT on February 27. Over that 55-hr interval the BMR pivots 70° counterclockwise.

#### *3.1.2. Example Emerging S-Shaped AR in the Northern Hemisphere*

Our example emerging S-shaped AR in the northern hemisphere is shown in Figure 2 and in its animation. The AR is number 30 (AR 12734) in Table 1. It emerges around latitude N08° in 2019 March.

Figure 2 and its animation show that the BMR pivots counterclockwise as it emerges and the AR becomes S-shaped in AIA 335 Å images, and that the BMR's two opposite-polarity sunspot clusters spread apart from each other as the BMR emerges. In the bottom panel of Figure 2, the time plots of the BMR's tilt and magnetic flux together show that the BMR completes its pivot after it completes its emergence. The BMR's first measured tilt is at 16:00 UT on March 5 and its last measured tilt is at 06:00 UT on March 7. Over that 38-hr interval the BMR pivots 67° counterclockwise.

#### *3.1.3. Example Emerging Z-shaped AR in the Northern Hemisphere*

Our example emerging Z-shaped AR in the northern hemisphere is shown in Figure 3 and its animation. The AR is number 34 (AR 12929) in Table 1. It emerges around latitude N08° in 2022 January, in sunspot cycle 25. It has negative leading polarity, the opposite of the positive leading polarity of the large majority of northern-hemisphere ARs in cycle 25.

Figure 3 and its animation show that the BMR pivots clockwise as it emerges and the AR becomes Z-shaped in AIA 335 Å images, and that the BMR's two opposite-polarity sunspot clusters spread apart from each other as the BMR emerges. In the bottom panel of Figure 3, the time plots of the BMR's tilt and magnetic flux together show that the BMR is still gradually emerging as its pivot gradually ends. The BMR's first measured tilt is at 21:00 UT on January 13 and its last measured tilt is at 19:00 UT on January 16. Over that 70-hr interval the BMR pivots 44° clockwise.

#### *3.1.4. Example Emerging Z-shaped AR in the Southern Hemisphere*

Our example emerging Z-shaped AR in the southern hemisphere is shown in Figure 4 and in its animation. The AR is number 37 (AR 13164) in Table 1. It emerges around latitude S18° in 2022 December.

Figure 4 and its animation show that the BMR pivots clockwise as it emerges and the AR becomes Z-shaped in AIA 335 Å images, and that the BMR's two opposite-polarity sunspot clusters spread apart from each other as the BMR emerges. In the bottom panel of Figure 4, the time plots of the BMR's tilt and magnetic flux together show that the BMR completes its pivot as it completes its emergence. The BMR's first measured tilt is at 04:00 UT on December 11 and its last measured tilt is at 18:00 UT on December 12. Over that 38-hr interval, the BMR pivots 57° clockwise.

### *3.2. Table 1*

For each of our 42 emerging ARs, the main results and other specifics are listed in Table 1. The first two columns of Table 1 give each AR's chronological ordinal number and NOAA AR number. The third column gives each AR's nominal solar latitude. Columns 3 – 6 give the date, time, and place on the solar disk at which each AR's emerging BMR first has at least one sunspot or sunspot pore in each of its two opposite polarity domains. That time is the time at which the emerging BMR's tilt angle is first measured. Column 7 gives which sigmoidal shape the AR acquires as it emerges. Column 8 gives the direction of pivot of each AR's emerging BMR. Columns 9 and 10 give each emerging BMR's first and last measured tilt angle. Column 11 gives the time between each BMR's first and last measured tilt angle. The last column, column 12, gives each BMR's amount of pivot, the angular sweep of the BMR's tilt from the first measured tilt angle to the last.

The most important result seen in Table 1 is the following. In every emerging S-shaped sigmoidal AR the emerging BMR pivots counterclockwise, and in every emerging Z-shaped sigmoidal AR the emerging BMR pivots clockwise.

Figure 5 is a histogram of the pivot amounts of the BMRs of the 42 emerging sigmoidal ARs in Table 1. In this histogram, each red square is for an emerging S-shaped AR in which the emerging BMR's pivot amount is in the pivot-amount range of that square, and each blue square is for an emerging Z-shaped AR in which the emerging BMR's pivot amount is in the pivot-amount range of that square. The average pivot amount of the 42 emerging BMRs is 35.3°.

## 4. Interpretation

Figure 6 is a progression of four schematic drawings depicting that counterclockwise pivot of the emerging BMR in the birth of an S-shaped sigmoidal active region is consistent with the AR's emerging Ω-loop flux rope having right-handed twist about the flux rope's center line.

In the top left panel is a drawing of an east-west Ω-loop flux rope that is viewed horizontally from the south. The Ω loop is buoyantly rising to the top of the convection zone and will soon begin emerging through the photosphere. We have tailored this Ω-loop flux rope for its emergence to qualitatively fit the observed emergence of our first example emerging S-shaped sigmoidal AR, AR 11680 shown in Figure 1. The polarity of the leading (west) leg of the Ω loop is positive, and the twist in the flux-rope field is right-handed. The twist pitch angle (relative to the flux rope's central field line) of a field line on the flux-rope's surface is drawn to be 45°. That is about 25° less steep than what the observations plausibly suggest for the Ω-loop flux rope that emerges to make AR 116980. We assume that the flux-rope field's twist pitch decreases inward monotonically to zero at flux rope's central field line.

The other three panels of Figure 6 are a sequence of schematic drawings of the emerging BMR and the emerging AR's coronal magnetic field viewed vertically from above. In the top right panel, only magnetic field that was very near the surface of the flux rope has emerged. Consequently, the tilt of the emerging BMR is plausibly 45° clockwise of east-west, and the emerging AR's coronal field shows no noticeable overall S-shaped twist. In the bottom left panel, field that was farther inside the flux rope (but not as far inside as the central field line) has emerged. That has caused the growing BMR to pivot counterclockwise to become less tilted to the pre-emerged Ω loops east-west direction. It has also caused the emerging AR's coronal field to begin to have discernible overall S-shaped (right-handed) twist. Finally, in the bottom right panel, about half of the field that was in the top of the pre-emerged Ω loop has emerged, the Ω loop's emergence is ending, the emerged BMR is directed east-west, and the AR's coronal magnetic field is now more obviously S-shaped.

If the magnetic field were the mirror-image of that depicted in Figure 6, i.e., if instead of the right-handed twist drawn in Figure 10, the Ω-loop flux rope's field had a comparable amount of left-handed twist and the field's twist pitch angle similarly decreased to zero at the flux rope's central field line, the emergence would produce a BMR that pivots clockwise and a sigmoidal AR that has Z-shaped coronal magnetic field.

As is seen in the AIA 335 Å images in Figures 1-4, and in AIA 335 Å images of all of our 42 active regions, there is much internal substructure in the form of coronal loops that – when the active region has become overall sigmoidal – are each shorter than the overall sigmoid, but, in aggregate, give the active region its overall sigmoidal shape. The schematic drawings in Figure 6 ignore the sigmoid's internal loop substructure. MHD simulations of the emergence of the top of an Ω loop from below the photosphere show that the rising top of the Ω loop becomes flattened as it reaches the photosphere. In these simulations, throughout the interior of the emerging BMR the convection in and below the photosphere concentrates the emerging magnetic flux into clumps ranging in scale from that of photospheric granules to that of supergranule convection cells (M. C. M. Cheung & H. Isobe 2014). We suppose that such flux clumping is the primary cause of the loop substructure seen in AIA 335 Å images of our sigmoidal active regions. Even though the emergence of an active-region flux-rope Ω loop has this evident complexity, we infer, from both our observations and the MHD simulations of the emergence of sigmoidal active regions cited in the Introduction, that the coronal magnetic field of the emerged sigmoidal active region retains the direction (handedness) of overall twist that the pre-emergent Ω-loop field has.

## 5. Summary and Discussion

Our search through 12 years of full-disk AIA 335 Å coronal EUV images, yielded only 42 ARs that each both emerges on the face of the Sun and becomes sigmoidal as it emerges on the face of the Sun. We tracked the emergence of each AR's BMR in HMI magnetograms and HMI continuum images. We measured each BMR's tilt angle to east-west at many times throughout the interval of the emergence in which the BMR had a sunspot cluster in each of its two opposite-polarity flux domains. The tilt angle that we measure is the tilt angle of the straight line through the visually located centroid of each of the BMR's two opposite-polarity sunspot clusters. From each emerging BMR's time sequence of measured tilt angles, we certify the direction (clockwise or counterclockwise) of BMR's pivot, and obtain the BMR's pivot amount, the net pivot in that direction from early to late in the BMR's emergence. The 42 BMR's range in pivot amount from about 10° to about 90°, and the average pivot amount is 35°.

A sigmoidal AR's coronal magnetic field either has enough net right-handed twist to make the AR noticeably S-shaped in coronal EUV images and coronal X-ray images, or has enough net left-handed twist to make the AR noticeably Z-shaped in such images. In our set of 42 randomly selected emerging sigmoidal ARs, the coronal magnetic field becomes S-shaped in 19 ARs and becomes Z-shaped in 23 ARs. For each S-shaped AR, the emerging BMR pivots counterclockwise, and for each Z-shaped AR, the emerging BMR pivots clockwise. This observed unanimous behavior matches what has been found in MHD simulations of the production of sigmoidal ARs by the emergence of an Ω-loop flux rope from below the photosphere into the corona: If the initially horizontal subsurface flux rope's magnetic field has right-handed twist, the emerging BMR pivots counterclockwise and the emerging AR's coronal field becomes S-shaped, and if the initial flux-rope's field has left-handed twist, the emerging BMR pivots clockwise and the emerging AR's coronal field becomes Z-shaped.

Our observational confirmation of the above behavior in MHD simulations of an Ω loop emerging from an initially horizontal subsurface twisted flux-rope field justifies the underlying premise of the MHD simulations, namely that a single-BMR sigmoidal AR emerges from an initially horizontal twisted-field flux rope. In our opinion, the unanimous agreement of our 42 observed emerging sigmoidal ARs with the MHD simulations also strengthens the tenet that every emerging BMR is made by an emerging flux-rope Ω loop.

In our search through 12 years of full-disk HMI magnetograms, HMI continuum images, and AIA 335 Å images we noticed that only a small minority of single-BMR ARs that emerge on the solar disk become noticeably sigmoidal in AIA 335 Å images. That characteristic implies that the magnetic field in the flux-rope Ω loop for most single-BMR ARs has less net twist than in the flux-rope Ω loops for single-BMR ARs that are noticeably sigmoidal in AIA 131 Å images. That implication raises the question of what determines the degree of net magnetic twist in an Ω-loop's flux rope before the Ω loop emerges to make a single-BMR AR.

Following R. L. Moore, S. K. Tiwari, N. K. Panesar, & A. C. Sterling 2020, we suppose that the initial flux rope is the twisted magnetic field that fills the Ω loop that will soon emerge, as in the top left panel of Figure 6. In this picture, the initial flux rope is roughly horizontal only near the top of the Ω loop. The rest of the flux rope fills the Ω loop's two legs, of which, a la R. L. Moore, S. K. Tiwari, N. K. Panesar, & A. C. Sterling 2020, we suppose the vertical extents sit in down flows at opposite edges of a convection cell. In this picture, at the end of the Ω loop's emergence, the centroid of one of the BMR's two opposite-polarity flux domains is roughly centered on one down flow and the centroid of the other flux domain is roughly centered on the other down flow, as in the bottom right panel of Figure 6. If that is the case, then we suppose that the twist in the Ω-loop flux-rope's field comes from the field in the legs being twisted by cyclonic spiraling of each down flow. Hence, we suppose that in only a small minority of Ω-loops does the spiraling down flow give the Ω-loop field enough net twist to make the emerged AR noticeably sigmoidal in coronal images.

The Coriolis force due to the Sun's rotation acts to make the down flows at the edges of convection cells spiral counterclockwise in the Sun's northern hemisphere and clockwise in the southern hemisphere**,** for convection cells as large or larger in diameter than supergranules (> ~ 30,000 km, T. L. Duvall & L. Gizon 2000; A. A. Pevtsov 2016). For each of our emerged active regions, the span of the BMR exceeds 30,000 km. Counterclockwise twisting of each of the Ω loop's legs gives the field in the top of the Ω loop left-handed twist, and clockwise twisting of each of the Ω loop's legs gives the field in the top of the Ω loop right-handed twist. For our picture, that is plausibly why most sigmoidal single-BMR ARs in the northern hemisphere are Z-shaped (have left-handed magnetic twist) and most sigmoidal single-BMR ARs in the southern hemisphere are S-shaped (have right-handed magnetic twist).

We assume that because the flows in the free convection in the convection zone are turbulent, a minority of down flows at the edges of convection cells spiral in the direction opposite that expected from the direction of the Sun's rotation (e.g., D. W. Longcope, G. H. Fisher, & A. A. Pevtsov 1998). For our picture, that is plausibly why a minority (6) of our 19 S-shape ARs are in the northern hemisphere, and a minority (4) of our 23 Z-shaped AR's are in the southern hemisphere.

Our emerging sigmoidal AR number 36 (AR 13029) emerges in the southern hemisphere at latitude S18°, pivots clockwise, and becomes Z-shaped in AIA 335 Å images, all of which is normal. This AR is exceptional in that its pivot amount exceeds 90°. Its measured pivot amount is 94.2° ± 1.4°. Pivot amounts more than 90° cannot be explained by only the emerging flux-rope field's twist about the flux rope's central field line. For this AR we suppose that left-handed twist in the pre-emergent flux-rope field is great enough that the flux rope has started to kink, i.e., that some of the twist in flux-rope field has gone into left-handed writhe of the flux rope. We suppose that the pre-emergent flux rope has enough initial left-handed writhe to make the clockwise pivot amount exceed 90° and be the measured amount.

Some studies of overall magnetic twist of active regions have found correlation of the overall magnetic twist with the tilt of the active regions from solar east-west (R. C. Canfield & A. A. Pevstov 1998; L. Tian, S. Bao, H. Zhang, & H. Wang 2001; L. Tian, D. Alexander, Y. Liu, & J. Yang 2005). That motivated us to look for correlation of tilt angle with pivot-amount angle at the end of emergence in our set of 42 single-BMR sigmoidal active regions. The pivot-amount angle is a measure of the direction and amount of overall twist in the active region's at the end of emergence, and is taken to be positive for counterclockwise pivot, and negative for clockwise pivot. The scatter plot of end-of-pivot tilt versus pivot-amount angle (Figure 7) shows an insignificant slight negative correlation of the two angles. That is, we find no significant correlation between the tilt of the active regions and their overall magnetic twist at the end of emergence.

D. W. Longcope & B. T. Welsch (2000) have pointed out that the overall twist in an active region's coronal magnetic field may continue to increase after the active region's flux emergence has ended. The continued increase plausibly comes from twist in the subphotospheric roots of the active region's magnetic field propagating up into the coronal field via Alfvenic magnetic twist waves. Evidence for this effect in newly emerged active regions was found by A. A. Pevtsov, V. M. Maleev, & D. W. Longcope (2003). Perhaps this effect is the reason that in the case of our example active region shown in Figure 2, the BMR continues to pivot for many hours after the end of the BMR's flux emergence.

Our observations do not prove that the scenario depicted in Figure 6 is the main way that our active regions become sigmoidal as the active region emerges and the active-region's BMR pivots, but we think that the agreement of our observations with MHD simulations of the birth of sigmoidal active regions strongly suggests that scenario. Pevtsov (2002) reviews other possible mechanisms for making an active region become sigmoidal, including shearing the BMR by shear flows in and below the photosphere. While such shearing and other mechanisms such as the upward propagation of magnetic twist from below the photosphere as proposed by D. W. Longcope & B. T. Welsch (2000) perhaps contribute to making some of our active regions become sigmoidal as they emerge, we think that any mechanism/process other than that depicted in Figure 6 is secondary. We think that the scenario depicted in Figure 6 is the most viable scenario for explaining our main finding: in the birth of each of our 42 single-BMR sigmoidal active regions – regardless of whether the active region emerges north or south of the equator – if the emerging BMR pivots counterclockwise the active region becomes S-shape sigmoidal, and if the emerging BMR pivots clockwise the active region becomes Z-shape sigmoidal.

The reviewer's comments helped us make the paper clearer and more complete. This work was supported by the Heliophysics Division of NASA's Science Mission Directorate through the Heliophysics Guest Investigators (HGI) program and the Heliophysics Supporting Research (HSR) program. SKT, RLM, VA, and NKP sincerely acknowledge support from NASA HGI grant (80NSSC21K0520), HSR grant (80NSSC23K0093) and/or NSF AAG award (no. 2307505). ACS and RLM acknowledge support from their NASA HSR grant. NKP acknowledges support from NASA's SDO/AIA grant (NNG04EA00C) and HSR grant (80NSS24K0258). We gratefully acknowledge the SDO/AIA and SDO/HMI science teams for providing the data used in this work. AIA and HMI are instruments on board the Solar Dynamics Observatory (SDO), a mission for NASA's Living With a Star program. We sincerely thank JSOC for their data distribution platform.

**References**

Babcock, H. W. 1961, ApJ, 133, 572

Canfield, R. C., Hudson, H. S., & McKenzie, D. E., 1999, GRL, 26, 627

Canfield, R. C., & Pevtsov, A. A. 1998, in Synoptic Solar Physics, ASP Conference Series, Vol. 140, Ed. K. S. Balasubramaniam, J. W. Harvey, & D. M. Rabin, 131

Charbonneau, P. 2020, LRSP, 17, 4

Cheung, M. C. M., & Isobe, H. 2014, LRSP, 113

Duvall, T. L., Jr., & Gizon, L. 2000, SoPh, 192, 177

Fan, Y. 2009, ApJ, 697, 1529

Fan Y. 2021, LRSP, 18, 5

Gibson, S. E., Fan, Y., Mandrini, C., Fisher, G., & Demoulin, P. 2004, ApJ, 617, 600

Green, L. M., Kleim, B., Torok, T., van Driel-Gesztelyi, L., & Attrill, G. D. R. 2007, SoPh, 246, 356

Howard, R. F. 1991, SoPh, 136, 251

Innes, D. E., & Teriaca, L. 2013, SoPh, 282, 453

Ishikawa, R., Tsuneta, S., & Jurcak, J. 2010, ApJ, 713, 1310

Lemen, J. R., Title, A. M., Akin, D. J., et al. SoPh, 275, 17

Liu, Z., Jiang, C., Feng, X., Zuo, P., & Wang, Y. 2023, ApJ Supp., 264, 13

Longcope, D. W., Fisher, G. H., & Pevtsov, A. A. 1998, ApJ, 507, 417

Longcope, D. W., & Welsch, B. T. 2000, ApJ, 545, 1089

Manchester, W., IV, Gombosi, T., DeZeeuw, D., & Fan, Y. 2004, ApJ. 610, 588

Miesch, M. S., Brun, A. S., DeRosa, M. L., & Toomre, J. 2008, ApJ, 673, 557

Moore, R. L., Cirtain, J. W., & Sterling, A. C. 2016, arXiv: 1606.05371

Moore, R. L., Falconer, D. A., Porter, J. G., & Suess, S. T. 1999, ApJ, 526, 505

Moore, R., & Rabin, D. 1985, ARA&A, 23, 239

Moore, R. L., Sterling, A. C., Cirtain, J. W., & Falconer, D. A. 2011, ApJ, 720, 757

Moore, R. L., Tiwari, S. K., Panesar, N. K., & Sterling, A. C. 2020, ApJL, 902, L35

Panesar, N. K., Sterling, A. C., Moore, R. L., et al 2019, ApJL, 887, L8

Panesar, N. K., Tiwari, S. K., Berghmans, D., Cheung, M. C. M., Muller, D., Auchere, F., & Zhukov, A. 2021, ApJL, 921, L20

Pesnell, W. D., Thompson, B. J., & Chamberlin, P. C. 2012, SoPh, 275, 3

Pevtsov, A. A. 2002, in Multi-Wavelength Observations of Coronal Structure and Dynamics, ed. P. C. H. Martens & P. Cauffman (Amsterdam, Pegamon), 125

Pevtsov, A. A. 2016, Ge&Ae, 56, 982

Pevtsov, A. A., Canfield, R. C., & Metcalf, T. R. 1995, ApJL, 440, L109

Pevtsov, A. A., Maleev, V. M., & Longcope, D. W. 2003, ApJ, 593, 1217

Rabin, D., Moore, R., & Hagyard, M. J. 1984, ApJ, 287, 404

Rust, D. M., & Kumar, A. 1996, ApJ, 464, L199

Raouafi, N.-E., Georgoulis, M. K., Rust, D. M., & Bernasconi, P. N. 2010, ApJ, 718, 981

Samanta, T., Tian, H., Yurchyshyn, V., et al 2019, Sci, 366, 890

Scherrer, P. H., Schou, J., Bush, R. I., et al. 2012, SoPh, 275, 207

Shibata, K., Nakamura, T., Matsumotos, T., et al 2007, Sci, 318, 1591

Spruit, H. C. 2011, in Theories of the Solar Cycle: A Critical View, Vol. 4, ed. M. P. Miralles & J. Sanchez Almeida (Berlin, Springer), 39

Stein, R. F., & Nordlund, A. 2012, ApJL, 753, L13

Sterling, A. C., Moore, R. L., Falconer, D. A., & Adams, M. 2015, Natur, 523, 437

Tian, L., Alexander, D., Yang, L., & Yang, J. 2005, SoPh, 229, 63

Tian, L., Bao, S., Zhang, H., & Wang, H. 2001, A&A, 374, 294

Tiwari, S. K., Panesar, N. K., Moore, R. L., et al 2019, ApJ, 887, 56

Vaiana, G. S., & Rosner, R. 1978, ARA&A, 16, 393

van Driel-Gesztelyi, L., & Green, L. M. 2015, LRSP, 249, 167

Wang, Y. M., & Sheeley, N. R., Jr. 1989, SoPh, 124, 81

Withbroe, G. L., & Noyes, R. W. 1977, ARA&A, 15, 363

Zirin, H. 1988, Astrophysics of the Sun, (Cambridge: Cambridge University Press)

Zwaan, C. 1978, ARA&A, 25, 83

**Table 1. Specifics of 42 Emerging Sigmoidal Active Regions**

| Ord No. | NOAA AR No. | Lat (deg) | Advent of Conjugate Pores | | | Sig-moid Shape (S or Z) | Pivot Direction (clockwise or counterclockwise) | Measured Angle to X-Axis | | | Pivot Amount (degrees) |
|---|---|---|---|---|---|---|---|---|---|---|---|
| | | | Date (YYYY/MM/DD) | Time (UT) | Place (x", y") | | | First (degrees) | Last (degrees) | Δt (hrs) | |
| 1 | 11295 | N21 | 2011/09/13 | 22:00 | 350, 188 | Z | clockwise | 69.6 ± 0.4 | 1.7 ± 0.3 | 26 | 67.9 ± 0.5 |
| 2 | 11318 | N21 | 2011/10/12 | 00:00 | -144, 238 | S | counterclockwise | -78.1 ± 0.8 | -24.2 ± 0.5 | 60 | 53..9 ± 0.9 |
| 3 | 11444 | N19 | 2012/03/22 | 18:00 | -745, 390 | Z | clockwise | 64.4 ± 0.2 | 19.4 ± 0.6 | 54 | 45.0 ± 0.6 |
| 4 | 11449 | S18 | 2012/03/28 | 10:00 | -62, -207 | S | counterclockwise | -9.6 ± 0.7 | 6.6 ± 0.6 | 32 | 16.2 ± 0.9 |
| 5 | 11460 | N16 | 2012/04/16 | 18:00 | -650, 308 | S | counterclockwise | -46.3 ± 0.5 | -3.6 ± 0.2 | 54 | 42.7 ± 0.5 |
| 6 | 11511 | N15 | 2012/06/21 | 14:00 | -63, 221 | Z | clockwise | 23.6 ± 0.8 | -32.1 ± 0.8 | 28 | 55.7 ± 1.1 |
| 7 | 11560 | N03 | 2012/08/29 | 10:00 | -668, -36 | Z | clockwise | 72.5 ± 1.0 | 24.8 ± 0.5 | 62 | 47.7 ± 1.1 |
| 8 | 11565 | N10 | 2012/09/03 | 04:00 | -495, 76 | Z | clockwise | 31.7 ± 0.8 | 7.7 ± 0.4 | 35 | 24.1 ± 0.9 |
| 9 | 11600 | N09 | 2012/10/26 | 04:00 | 28, 73 | Z | clockwise | -16.2 ± 1.1 | -24.6 ± 0.4 | 50 | 8.4 ± 1.2 |
| 10 | 11631 | N20 | 2012/12/11 | 21:00 | -151, 334 | Z | clockwise | 16.5 ± 1.6 | -19.8 ± 0.2 | 51 | 27.3 ± 1.6 |
| 11 | 11645 | S13 | 2013/01/02 | 20:00 | -200, -173 | S | counterclockwise | -32.2 ± 1.6 | 3.8 ± 0.9 | 34 | 36.0 ± 1.8 |
| 12 | 11678 | N10 | 2013/02/18 | 06:00 | 265, 278 | Z | clockwise | 18.6 ± 3.3 | -4.2 ± 0.4 | 48 | 22.8 ± 3.3 |
| 13 | 11680 | S28 | 2013/02/24 | 17:00 | -680, -398 | S | counterclockwise | -63.9 ± 0.4 | 6.0 ± 0.2 | 55 | 69.9 ± 0.4 |
| 14 | 11682 | S18 | 2013/02/24 | 06:00 | -320, -190 | S | counterclockwise | -16.4 ± 1.7 | -4.8 ±0.3 | 42 | 11.6 ± 1.7 |
| 15 | 11707 | S11 | 2013/03/28 | 12:00 | -589, -92 | S | counterclockwise | -33.9 ± 1.2 | 15.0 ± 0.4 | 21 | 48.9 ± 1.3 |
| 16 | 11910 | N01 | 2013/11/27 | 17:00 | -86, -4 | S | counterclockwise | -18.0 ± 0.8 | 5.8 ± 0.1 | 31 | 23.8 ± 0.8 |
| 17 | 11992 | S19 | 2014/02/26 | 00:00 | -525, -245 | S | counterclockwise | 3.5 ± 1.7 | 42.9 ±0.7 | 36 | 38.4 ± 1.8 |
| 18 | 12003 | N06 | 2014/03/10 | 10:00 | 39, 232 | Z | clockwise | 26.4 ± 2.0 | -6.7 ± 0.2 | 26 | 33.1 ± 2.0 |
| 19 | 12008 | S11 | 2014/03/18 | 21:00 | -372, -60 | S | counterclockwise | -52.0 ± 0.8 | 4.9 ± 0.5 | 57 | 56.9 ± 0.9 |
| 20 | 12036 | S18 | 2014/04/13 | 13:00 | -472, -213 | S | counterclockwise | -16.4 ± 0.8 | 10.1 ± 0.5 | 89 | 26.6 ± 0.9 |
| 21 | 12082 | N17 | 2014/06/05 | 02:00 | -722, 282 | Z | clockwise | 33.8 ± 2.5 | -25.4 ± 0.5 | 64 | 59.2 ± 2.5 |
| 22 | 12189 | N21 | 2014/10/14 | 12:00 | -386, 258 | Z | clockwise | -20.2 ± 0.3 | -48.3 ± 0.4 | 24 | 28.1 ± 0.5 |
| 23 | 12234 | N05 | 2014/12/09 | 20:00 | -551, 82 | Z | clockwise | 39.0 ± 0.9 | -5.1 ± 0.2 | 52 | 44.1 ± 0.9 |
| 24 | 12273 | S03 | 2015/01/26 | 09:00 | -191, 41 | S | counterclockwise | -32.8 ± 0.6 | -11.7 ± 0.2 | 45 | 21.1 ± 0.6 |
| 25 | 12440 | N19 | 2015/10/26 | 11:00 | -79, 242 | S | counterclockwise | -51.6 ± 0.9 | -29.9 ± 0.3 | 43 | 21.7 ± 0.9 |
| 26 | 12441 | N14 | 2015/10/27 | 06:00 | -795, 183 | Z | clockwise | 6.5 ± 1.2 | -6.5 ± 0.6 | 54 | 13.0 ± 1.3 |
| 27 | 12460 | N12 | 2015/11/25 | 16:00 | -511, 189 | Z | clockwise | 28.6 ± 0.5 | -22.4 ± 0.4 | 26 | 51.0 ± 0.6 |
| 28 | 12579 | N12 | 2016/08/22 | 06:00 | -442, 90 | Z | clockwise | 31.2 ± 0.4 | 21.6 ± 0.4 | 42 | 9.6 ± 0.6 |

| Table 1. Specifics of 42 Emerging Sigmoidal Active Regions (Continued) | | | | | | | | | | | |
|---|---|---|---|---|---|---|---|---|---|---|---|
| Ord No. | NOAA AR No. | Lat (deg) | Advent of Conjugate Pores | | | Sig-moid Shape (S or Z) | Pivot Direction (clockwise or counterclockwise) | Measured Angle to X-Axis | | | Pivot Amount (degrees) |
| | | | Date (YYYY/MM/DD) | Time (UT) | Place (x", y") | | | First (degrees) | Last (degrees) | $\Delta t$ (hrs) | |
| 29 | 12696 | S11 | 2018/01/15 | 02:00 | -590 -141 | S | counterclockwise | -107.3 ± 0.7 | -49.3 ± 0.9 | 56 | 58.0 ± 1.1 |
| 30 | 12734 | N08 | 2019/03/05 | 16:00 | -480, 256 | S | counterclockwise | -87.9 ± 0.9 | -21.1 ± 0.5 | 38 | 66.8 ± 1.0 |
| 31 | 12805 | S23 | 2021/02/22 | 12:00 | -117, -266 | S | counterclockwise | -36.7 ± 1.3 | 6.3 ± 0.6 | 54 | 43.0 ± 1.4 |
| 32 | 12858 | N12 | 2021/08/18 | 10:00 | -128, 101 | Z | clockwise | 36.6 ± 1.3 | -7.4 ± 0.3 | 32 | 44.0 ± 1.3 |
| 33 | 12873 | N26 | 2021/09/20 | 04:00 | -60, 315 | Z | clockwise | 19.4 ± 1.1 | -14.0 ± 0.4 | 38 | 33.4 ± 1.2 |
| 34 | 12929 | N08 | 2022/01/13 | 21:00 | -191, 196 | Z | clockwise | 76.2 ± 3.0 | 32.3 ± 0.2 | 70 | 43.9 ± 3.0 |
| 35 | 12933 | S21 | 2022/01/15 | 22:00 | -311, -284 | Z | clockwise | 82.4 ± 1.8 | 29.8 ± 0.2 | 38 | 52.6 ± 1.8 |
| 36 | 13029 | S18 | 2022/06/08 | 19:00 | -70, -284 | Z | clockwise | 138 ± 1.3 | 43.8 ± 0.5 | 41 | 94.2 ± 1.4 |
| 37 | 13164 | S18 | 2022/12/11 | 04:00 | 320, -323 | Z | clockwise | 70.5 ± 0.4 | 13.9 ± 0.1 | 38 | 56.6 ± 0.4 |
| 38 | 13179 | N24 | 2022/12/29 | 06:00 | 4, 270 | S | counterclockwise | -23.2 ± 0.6 | -1.6 ± 0.2 | 60 | 21.6 ± 0.6 |
| 39 | 13315 | S17 | 2023/05/24 | 00:00 | -663, -268 | S | counterclockwise | -27.8 ± 0.7 | 2.0 ± 0.1 | 48 | 29.8 ± 0.7 |
| 40 | 13562 | S09 | 2024/01/29 | 11:00 | 101, -63 | S | counterclockwise | -33.1 ± 0.2 | 8.4 ± 0.4 | 37 | 41.5 ± 0.4 |
| 41 | 13597 | N07 | 2024/02/28 | 05:00 | -432, 220 | Z | clockwise | 90.8 ± 0.7 | 59.6 ± 0.1 | 31 | 31.2 ± 0.7 |
| 42 | 13606 | N08 | 2024/03/11 | 14:00 | -706, 215 | Z | clockwise | 21.0 ± 0.7 | -9.1 ± 0.3 | 52 | 30.1 ± 0.8 |

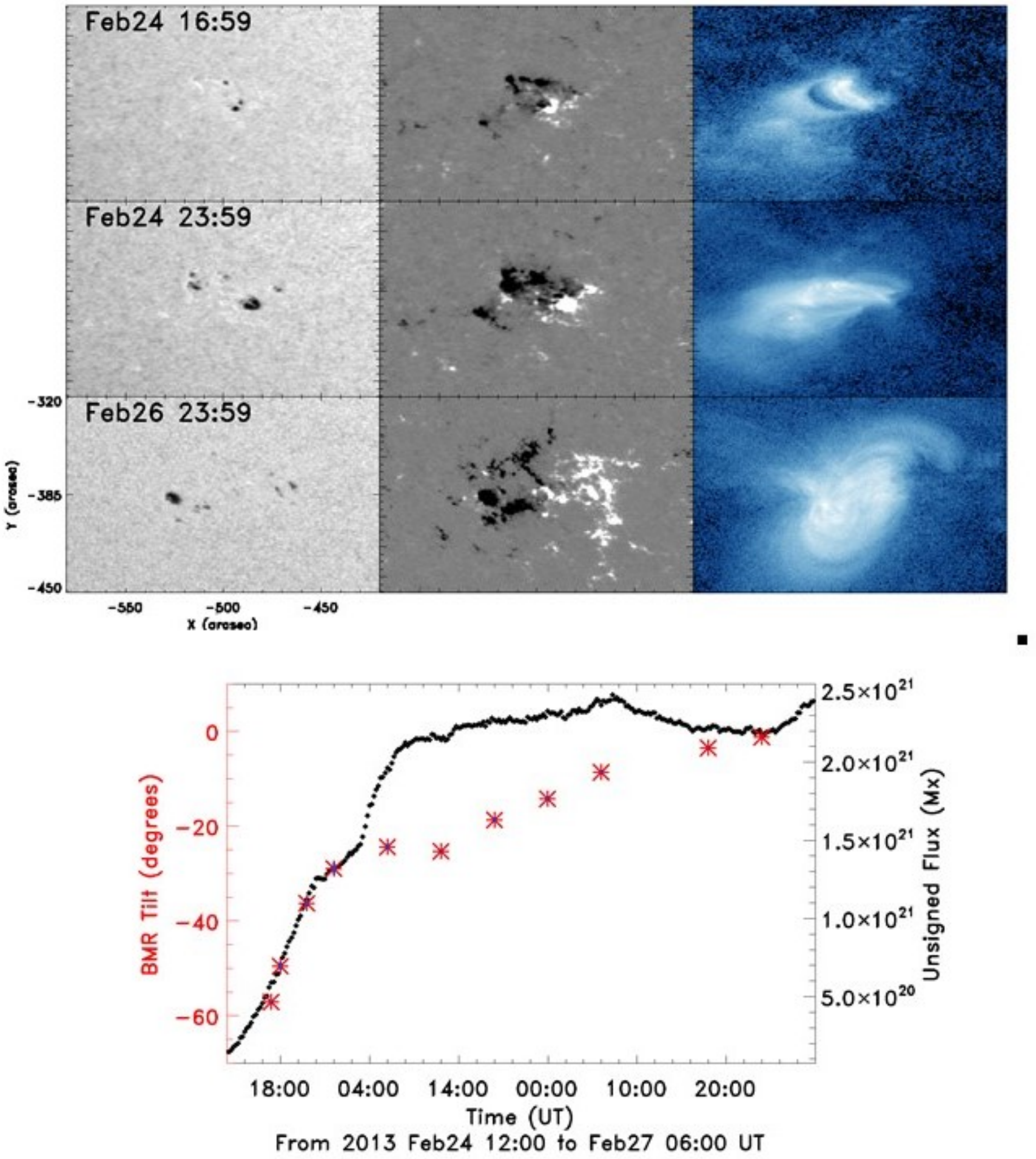

Figure 1. Evolution of sigmoidal AR number 13 (AR 11680) in the southern hemisphere at latitude S28°, from the start of emergence on 2013 February 24 to 06:00 UT on February 27. In each row, the panel on the left is an HMI continuum image, the middle panel is a simultaneous HMI magnetogram having the same field of view covering the AR, and the panel on the right is a simultaneous AIA 335 Å image having the same field of view. The time in each row is given in the row's continuum image. The first row is early in the emergence, when the BMR first has one or more discernible opposite-polarity sunspot pores, is tilted 64° clockwise of east-west, and the AR is not noticeably sigmoidal in AIA 335 Å images. The second row is seven hours later in the emergence, when the BMR is tilted 29° clockwise of east-west, and the AR is beginning to show right-handed overall magnetic shear and twist in AIA 335 Å images. The third row is two days after the second row, when emergence is finished, the BMR is tilted 6° counterclockwise of east-west, and the AR is definitely overall S-shaped in AIA 335 Å images. In the bottom panel are the time profile of the BMR's measured tilt angle and the time profile of the AR's measured unsigned magnetic flux. The uncertainty in each measured BMR tilt angle is smaller than the vertical span of the asterisk centered on the measured value. The animation spans the 66 hours spanned in the bottom panel, starts at 12:00 UT on 2013 February 24, ends at 00:60 UT on February 27, its cadence is 15 minutes, and its field of view is the same as in this figure.

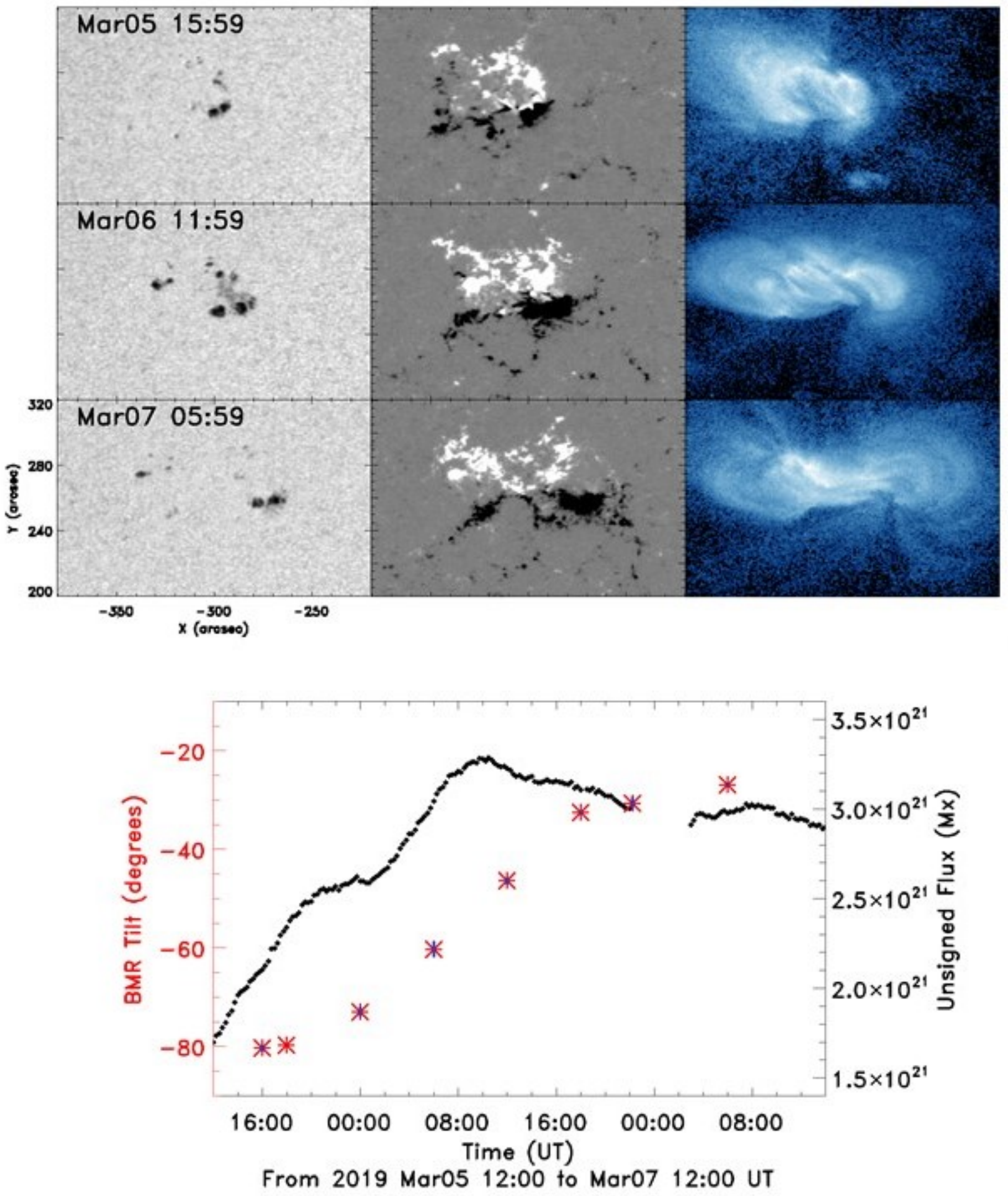

Figure 2. Evolution of sigmoidal AR number 30 (AR 12734) in the northern hemisphere at latitude N08°, from early in its emergence on 2019 March 5 to 12:00 UT on March 7. The layout is the same as for Figure 1. The first row is soon after the emerging BMR has sunspot pores in each of its two opposite-polarity flux domains. The BMR is now tilted 88° clockwise of east-west, and the AR is not noticeably sigmoidal in AIA 335 Å images. The second row is twenty hours later in the emergence than the first row. Now, the BMR is tilted 46° clockwise of east-west, and the AR shows right-handed overall magnetic shear and twist and is overall S-shaped in AIA 335 Å images. The third row is eighteen hours after the second row. Emergence has ended, the BMR is tilted 22° clockwise of east-west, and the AR remains overall S-shaped in AIA 335 Å images. In the bottom panel, the two time profiles show that the BMR is still completing its counterclockwise pivot after the AR's unsigned magnetic flux stops increasing. The uncertainty in each measured BMR tilt angle is smaller than the vertical span of the asterisk centered on the measured value. The animation spans the 48 hours spanned by the bottom panel, starts at 12:00 UT on 2019 March 5, ends at 12:00 UT on March 7, its cadence is 15 minutes, and its field of view is the same as in this figure. In the AIA 335 Å panel of the animation there is a short data gap in the second half of the animation.

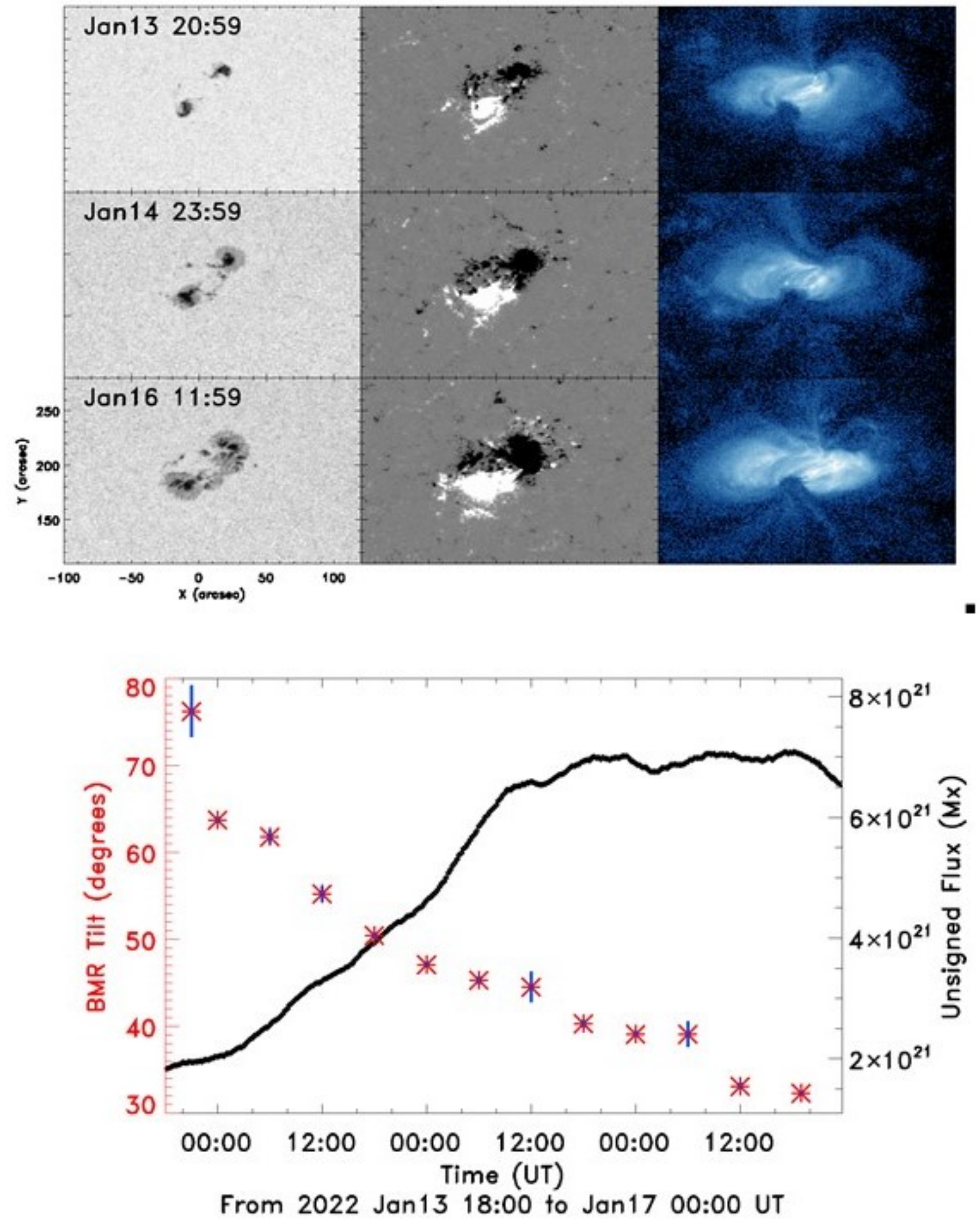

Figure 3. Evolution of sigmoidal AR number 34 (AR12929) in the northern hemisphere at latitude N08°, from near the start of emergence on 2022 January 13 to 00:00 UT on January 17. The layout is the same as for Figure 1. The first row is early in the emergence, when the BMR first has one or more discernible opposite-polarity sunspot pores, is tilted 76° counterclockwise of east-west, and the AR is not yet obviously sigmoidal in AIA 335 Å images. The second row is 27 hours after the first row. Now, the BMR is tilted 47° counterclockwise from east-west, and the AR shows overall left-handed magnetic shear and twist and is discernibly Z-shaped in AIA 335 Å images. The third row is 36 hours after the second row. Now, the BMR's pivot is ending while the emergence gradually continues, the BMR is tilted 33° counterclockwise of east-west, and the AR has become more starkly z-shaped in AIA 335 Å images. In the bottom panel the two time profiles show that the BMR's pivot is clockwise, and that the AR completes its emergence near when the BMR's pivot is ending. The uncertainty in each measured BMR tilt angle is shown by a blue error bar when the error bar's vertical span is comparable to or larger than the vertical span pf the asterisk centered on the measured value. The animation spans the 78 hours spanned in the bottom panel, starts at 18:00 on 2022 January 13, ends at 00:00 UT on January 17, its cadence is 15 minutes, and its field of view is the same as in this figure.

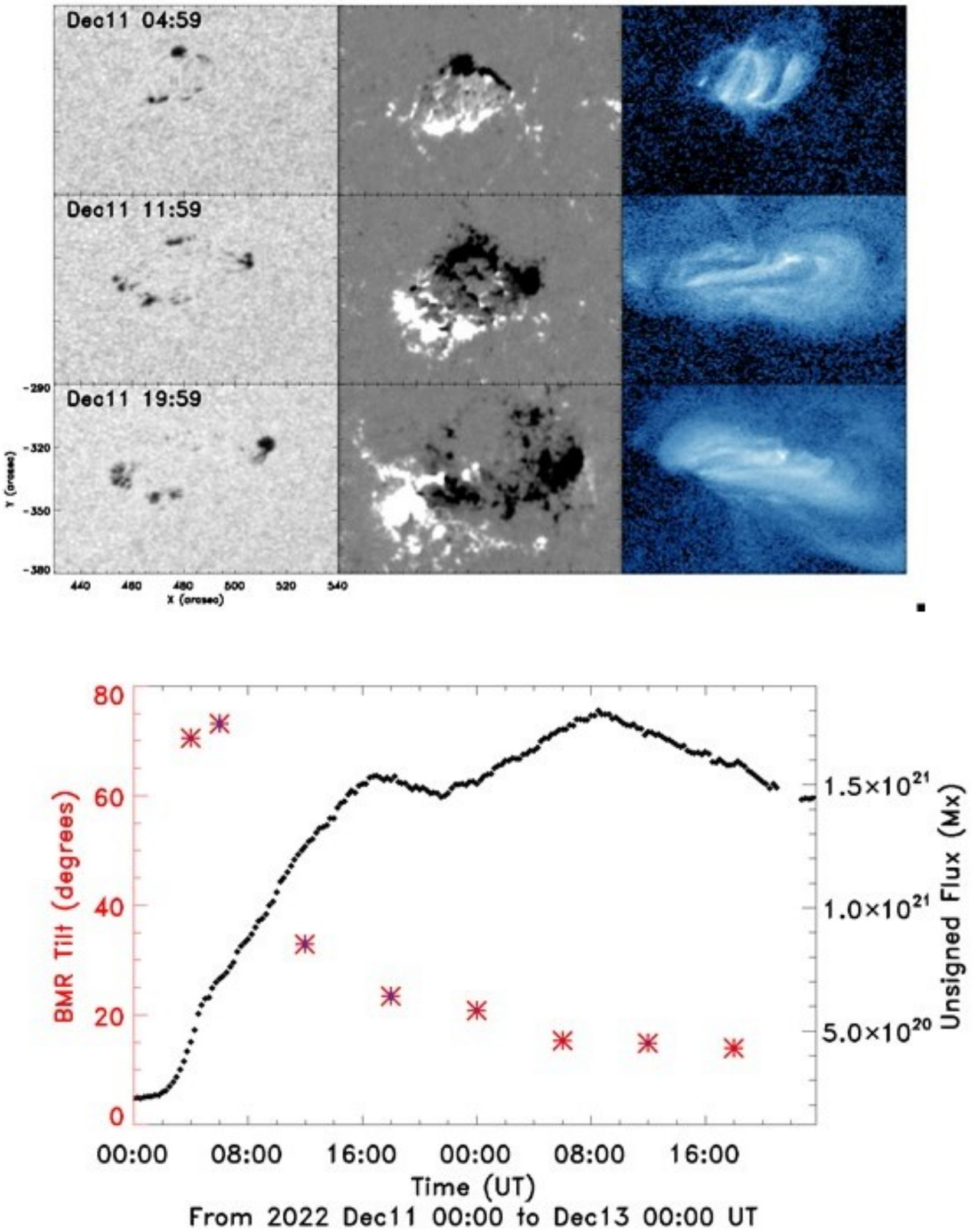

Figure 4. Evolution of sigmoidal AR number 37 (AR 13164) in the southern hemisphere at latitude S18°, from the start of emergence on 2022 December 11 to 00:00 UT on December 13. The layout is the same as for Figure 1. The first row is early in the emergence, roughly an hour after the BMR first has opposite-polarity sunspot pores. The BMR is tilted approximately 72° counterclockwise from east-west, and the AR is not noticeably sigmoidal in AIA 335 Å images. In the second row, seven hours after the first row, emergence is continuing, the BMR is tilted 34° counterclockwise of east-west, and the AR has become Z-shaped in AIA 335 Å images. In the third row, eight hours after the second row, emergence is ending, and the BMR is tilted approximately 22° counterclockwise of east-west, has completed all but a small fraction of its pivot, and the AR continues to be Z-shaped in AIA 335 Å images. In the bottom panel, the two time plots show that the pivot stops increasing near when the AR's magnetic flux stops increasing. The uncertainty in each measured BMR tilt angle is smaller than the vertical span of the asterisk centered on the measured value. The animation spans the 48 hours spanned in the bottom panel, starts as 00:00 UT on 2022 December 11, ends at 00:00 UT on December 13, its cadence is 15 minutes, and its field of view is the same as in this figure.

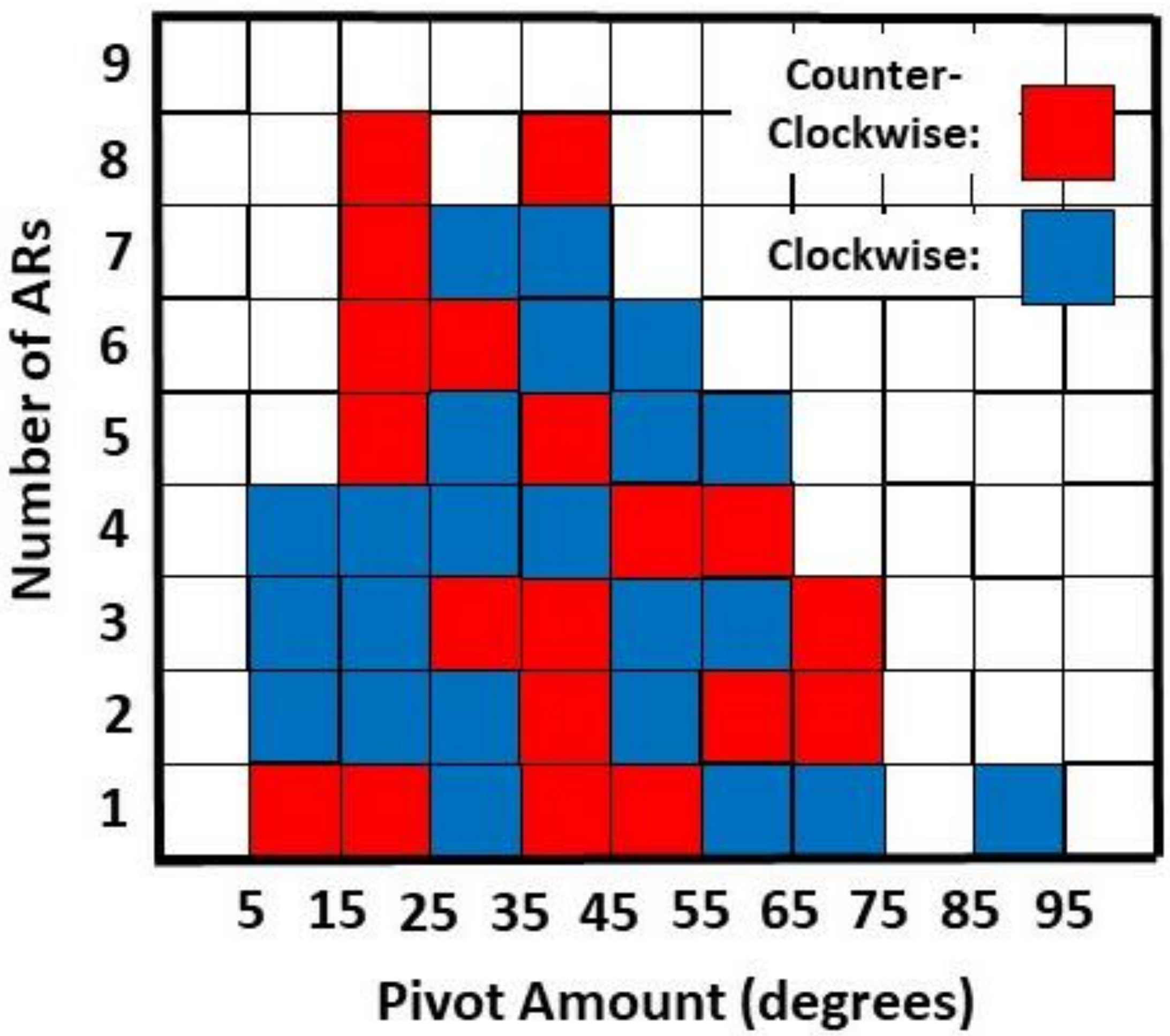

Figure 5. Histogram of the measured pivot amount for our 42 emerging sigmoidal ARs. Each colored square is for an AR that has a pivot amount in the pivot-amount range of that square. Each AR's measured pivot amount is listed in Table 1. Red squares are for ARs in which the emerging BMR pivots counterclockwise as the AR's coronal magnetic field becomes overall S-shaped. Blue squares are for ARs in which the emerging BMR pivots clockwise as the AR's coronal magnetic field becomes overall Z-shaped.

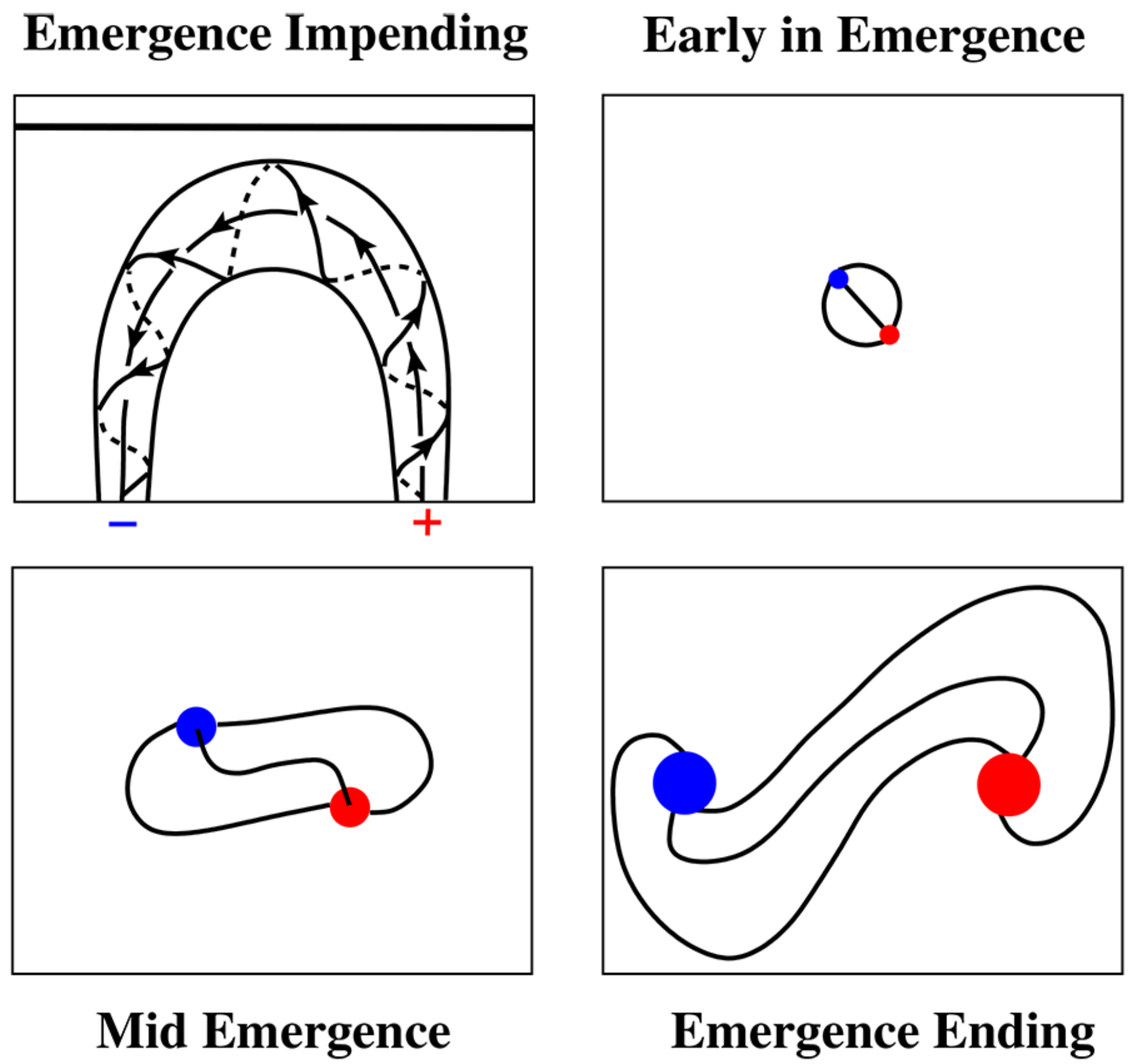

Figure 6. Four sequential schematic drawings depicting the production of an S-shaped sigmoidal AR by the emergence of an Ω-loop flux rope in which the magnetic field has right-handed twist about the flux rope's central field line. Top Left: Sketch of a buoyantly rising east-west pre-emergent Ω-loop flux rope viewed horizontally from the south. The thick black straight horizontal line close above the rising Ω loop is the Sun's photosphere surface. Only the outline of the flux-rope's surface, the flux rope's central field line, and a field line on the flux rope's surface are drawn. The pitch of the surface field's twist with respect to the central field line is drawn to be 45°. We assume the field's twist pitch angle monotonically decreases inward to zero at the central field line. We also assume that the Ω-loop flux rope's two opposite-polarity vertical legs each sit in a down flow at opposite edges of a convection cell. Top Right, Bottom Left, and Bottom Right: Three sequential sketches of the emerging Ω loop in the corona, viewed vertically from above. The growing red disk and blue disk represent the emerging BMR's growing positive and negative magnetic flux domains. The black lines connecting the two disks represent the emerging AR's coronal magnetic field lines. As the Ω-loop flux rope emerges through the photosphere into the corona, the BMR initially tilts 45° clockwise of east-west and then pivots counterclockwise to become directed east-west as its two growing opposite-polarity flux domains spread apart and the AR's coronal field becomes S-shaped. At the end of emergence, the BMR is directed east-west as in the Bottom Right panel, and each of the BMR's two opposite-polarity flux domains is centered on the down flow in which its leg of the Ω loop is trapped: Each is centered on the position of the vertical portion of its leg of the Ω loop in the Top Left panel, as comparison of those two panels shows.

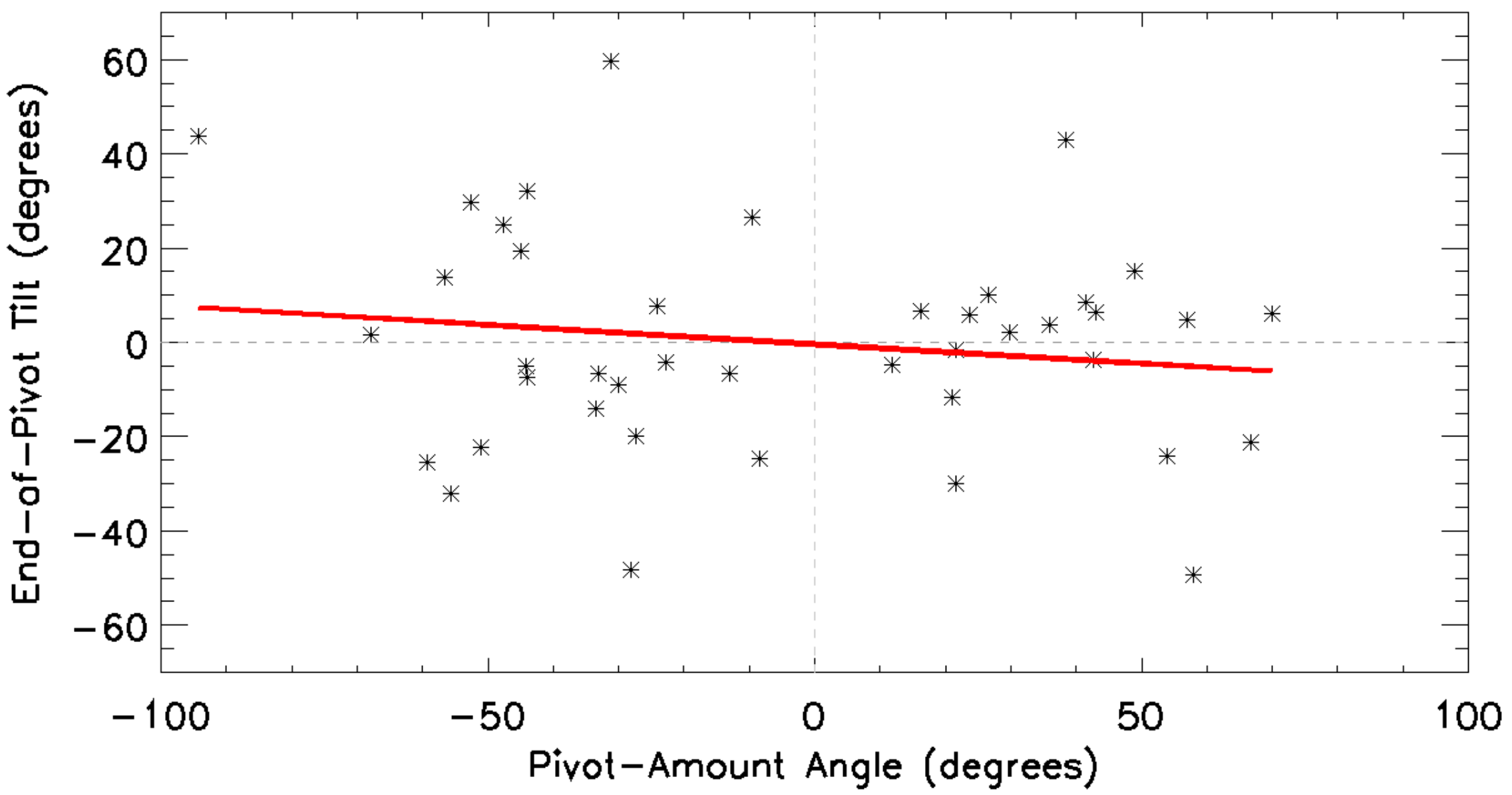

**Figure 7. Scatter plot of measured end-of-pivot tilt versus measured pivot-amount angle at the end of the emergence (both from Table 1) for our 42 single-BMR sigmoidal active regions. The pivot-amount angle is taken to be positive for counterclockwise pivot and negative for clockwise pivot. The red line is the linear least-squares fit to the 42 points. The slope and 1-σ uncertainty of the fit are: - 0.08 ± 0.08.**

**Animation Legends:**

Figure 1 animation. Left panel: HMI continuum images. Middle panel: HMI magnetograms. Right panel: AIA 335 Å images. The cadence is 15 minutes. The animation spans the 66 hours spanned by the time axis in the bottom panel of Figure 1. The animation starts at 12:00 UT on 2013 February 24, ends at 00:00 UT on February 27, and has a duration of about 30 seconds. The field of view is the same as in Figure1.

Figure 2 animation. The format and cadence are the same as for Figure 1 animation. The animation spans the 48 hours spanned by the time axis in the bottom panel of Figure 2. The animation starts at 12:00 UT on 2019 March 5, ends at 12:00 UT on March 7, and has a duration of about 20 seconds. The field of view is the same as in Figure 2.

Figure 3 animation. The format and cadence are the same as for Figure 1 animation. The animation spans the 78 hours spanned by the time axis in the bottom panel of Figure 3. The animation starts at 12:00 UT on 2022 January 13, ends at 00:00 UT on January 17, and has a duration of about 30 seconds. The field of view is the same as in Figure 3.

Figure 4 animation. The format and cadence are the same as for Figure1 animation. The animation spans the 48 hours spanned by the time axis in the bottom panel of Figure 4. The animation starts at 00:00 UT on 2022 December 11, ends at 00:00 UT on December 13, and has a duration of about 20 seconds. The field of view is the same as in Figure 4.